\documentclass[twocolumn]{aastex631}

\journalinfo{Accepted by The Astrophysical Journal Letters: September 1, 2022}

\shorttitle{Ultrafaint dwarfs in the M81 group}
\shortauthors{Bell et al.}

\begin{document}

\title{Ultrafaint Dwarf Galaxy Candidates in the M81 Group: Signatures of Group Accretion}

\correspondingauthor{Eric Bell}
\email{ericbell@umich.edu}

\author[0000-0002-5564-9873]{Eric F.\ Bell}
\affiliation{Department of Astronomy, University of Michigan, 1085 S. University Ave, Ann Arbor, MI 48109-1107, USA}

\author[0000-0003-2599-7524]{Adam Smercina}
\affiliation{Astronomy Department, University of Washington, Box 351580, U.W. Seattle, WA 98195-1580, USA}

\author[0000-0003-0511-0228]{Paul A.\ Price}
\affiliation{Department of Astrophysical Sciences, Princeton University, Princeton, NJ 08544, USA}

\author[0000-0001-9269-8167]{Richard D'Souza}
\affiliation{Vatican Observatory, Specola Vaticana, V-00120, Vatican City State}

\author[0000-0001-6380-010X]{Jeremy Bailin}
\affiliation{Department of Physics and Astronomy, University of Alabama, Box 870324, Tuscaloosa, AL 35487-0324, USA}

\author[0000-0001-6982-4081]{Roelof S.\ de Jong}
\affiliation{Leibniz-Institut f\"{u}r Astrophysik Potsdam (AIP), An der Sternwarte 16, D-14482 Potsdam, Germany}

\author[0000-0003-2294-4187]{Katya Gozman}
\affiliation{Department of Astronomy, University of Michigan, 1085 S. University Ave, Ann Arbor, MI 48109-1107, USA}

\author[0000-0002-2502-0070]{In Sung Jang}
\affiliation{Leibniz-Institut f\"{u}r Astrophysik Potsdam (AIP), An der Sternwarte 16, D-14482 Potsdam, Germany}
\affiliation{Department of Astronomy and Astrophysics, University of Chicago, Chicago, IL 60637, USA}

\author[0000-0003-2325-9616]{Antonela Monachesi}
\affiliation{Instituto de Investigaci\'{o}n Multidisciplinar en Ciencia y Tecnolog\'{\i}a, Universidad de La Serena, R\'{a}ul Bitr\'{a}n 1305, La Serena, Chile}
\affiliation{Departamento de Astronom\'{i}a, Universidad de La Serena, Av. Juan Cisternas 1200 N, La Serena, Chile}

\author[0000-0001-9852-9954]{Oleg Y.\ Gnedin}
\affiliation{Department of Astronomy, University of Michigan, 1085 S. University Ave, Ann Arbor, MI 48109-1107, USA}

\author[0000-0002-0558-0521]{Colin T.\ Slater}
\affiliation{Astronomy Department, University of Washington, Box 351580, U.W. Seattle, WA 98195-1580, USA}



\begin{abstract}
The faint and ultrafaint dwarf galaxies in the Local Group form the observational bedrock upon which our understanding of small-scale cosmology rests. In order to understand whether this insight generalizes, it is imperative to use resolved-star techniques to discover similarly faint satellites in nearby galaxy groups. We describe our search for ultrafaint galaxies in the M81 group using deep ground-based resolved-star data sets from Subaru’s Hyper Suprime-Cam. We present one new ultrafaint dwarf galaxy in the M81 group and identify five additional extremely low surface brightness candidate ultrafaint dwarfs that reach deep into the ultrafaint regime to $M_V \sim -6$ (similar to current limits for Andromeda satellites). These candidates' luminosities and sizes are similar to known Local Group dwarf galaxies Tucana B, Canes Venatici I, Hercules, and Bo\"otes I. Most of these candidates are likely to be real, based on tests of our techniques on blank fields. Intriguingly, all of these candidates are spatially clustered around NGC 3077, which is itself an M81 group satellite in an advanced state of tidal disruption. This is somewhat surprising, as M81 itself and its largest satellite M82 are both substantially more massive than NGC 3077 and by virtue of their greater masses, would have been expected to host as many or more ultrafaint candidates. These results lend considerable support to the idea that satellites of satellites are an important contribution to the growth of satellite populations around Milky Way--mass galaxies. 
\end{abstract}



\section{Introduction} \label{sec:intro}

As the lowest-mass galaxies, the numbers and properties of ultrafaint dwarfs (UFDs; $M_V>-7.7$; \citealt{Simon2019,DW2020}) are extremely sensitive to critical aspects of our theoretical understanding of galaxy formation \citep{Agertz2020} and are among the best existing probes of the nature of dark matter (DM; \citealt{Bullock2017,Nadler2021}).
Due to their low masses, small variations in galaxy formation physics result in orders-of-magnitude scatter in galaxy stellar-to-halo mass ratios \citep{Bullock2017,Fitts2017,Munshi2019,Agertz2020} --- a behavior that will manifest itself in the luminosity function and properties of UFD satellites \citep{Smercina2018,Bose2020,Carlsten2021}. 

Because UFDs are intrinsically faint and have extremely low surface brightnesses, the only way of discovering them has been by seeking concentrations of individual resolved stars in survey datasets \citep[e.g.,][]{belokurov2007,Koposov2015,DW2020}, limiting the most sensitive searches to the Milky Way \citep{DW2020} and M31 \citep{McConnachie2018}. There are signatures in both satellite systems of substructure and the accretion of satellites in groups: the delivery of satellites by the Magellanic Clouds \citep{Koposov2015,Patel2020}, differences between the star formation quenching times \citep{Weisz2019,DSouza2021} and radial profiles of the Milky Way and M31 satellites \citep{Samuel2020}, and claims of alignments or planes of satellites \citep{Pawlowski2012,Ibata2013}. Because the Milky Way and M31 experienced particular growth and accretion histories, our models --- which have been calibrated entirely in the Local Group by necessity --- may not accurately describe the satellite populations of a wider, more representative set of groups \citep[e.g.,][]{Carlsten2022,Smercina2022}.

While the survey power of the Vera C. Rubin Observatory and Nancy Grace Roman Space Telescope will spur rapid progress in this field, current facilities (e.g., Magellan's Megacam or Subaru's Hyper Suprime-Cam (HSC)) already allow the discovery of faint (e.g., \citealt{Smercina2017}, \citealt{Okamoto2019} in the M81 group) and ultrafaint (e.g., \citealt{Mutlu2022}, \citealt{Sand2022}) galaxies \edit1{in} the Local Volume ($D<5\,$Mpc). The M81 group ($D=3.6$\,Mpc; \citealt{RS11}) is particularly interesting to study. M81 has a rich satellite population; diffuse light searches with the Canada-France-Hawaii Telescope \citep{Chiboucas2013} and more recent resolved-star work \citep{Smercina2017,Okamoto2019} have revealed 17(!) new group members in the past decade. Deep multiband data \citep{Okamoto2015,Smercina2020} are available, allowing the development and testing of satellite search methods (see also \citealt{Mutlu2021}). Furthermore, M81 is undergoing a group-scale interaction involving the recent arrival and tidal disruption of at least two large satellites  \citep{yun1994,Okamoto2015,Smercina2020}, offering a possibility to study the impacts of satellite delivery in group accretions \citep{LiHelmi2008,Deason2015,DSouza2021}. 

Here, we report the discovery using resolved-star techniques of one new M81 group UFD, and we present five lower surface brightness candidate UFDs, with absolute magnitudes reaching toward $M_V \sim -6$. 

\section{Observations}
\label{sec:reduction}
We combine two datasets from HSC \citep{HSC}. \citet{Okamoto2015} surveyed the M81 group in $g$ and $i$ bands using 7 HSC pointings (each $\sim$1\fdg5  field of view), centered on M81. The four eastern pointings have excellent image quality with point-sources size of 0\farcs7--0\farcs9, and 50\% completeness limits $i\sim26.2$. The three western pointings have worse image quality, and while we analyze them, they do not have competitive depth for dwarf searches \citep{Okamoto2019}. \citet{Smercina2020} surveyed two pointings in each of three ($g,r,i$) filters, chosen to cover the outer regions of M81, M82, and NGC 3077. Image depth was nearly uniform across the two fields, yielding extinction-corrected point-source detection limits of $g\,{=}\,27$, $r\,{=}\,26.5$, and $i\,{=}\,26.2$, measured at $\sim$5$\sigma$. Seeing was relatively stable, resulting in consistent point-source sizes of 0\farcs7--0\farcs8.  

Both datasets were reduced with the HSC optical imaging pipeline, which is a fork of the Legacy Survey of Space and Time (LSST) pipeline whose main features are regularly re-integrated with the LSST pipeline \citep{Bosch2018}. The pipeline performs photometric and astrometric
calibration using the Pan-STARRS1 catalog \citep{Magnier2013}, reporting final magnitudes in the HSC natural system. Sources detected in $i$ band determine
reference positions for forced photometry, which is performed on co-added image stacks in all available passbands. All magnitudes were corrected for Galactic extinction following \cite{SFD} adopting the updated extinction coefficients from \cite{Schlafly2011}. For this work, we use size measurements derived using the \texttt{ext\_shapeHSM\_HsmSourceMoments} algorithms, which have been optimized for applications such as weak lensing where shape measurements are critical \citep{Hirata2003,Mandelbaum2005}.

In order to search for UFD satellites, we must differentiate between stars at the distance of the M81 group
and much more numerous unresolved background galaxies. There
is no single perfect method; consequently, we use three different methods.
\begin{enumerate}
  \item Morphology: In each passband, we determine the spatially variable point-spread function (PSF) on $0\fdg25$/$0\fdg4$ scales (in the \citealt{Okamoto2015} and \citealt{Smercina2020} data sets respectively) using bright stars $19<i<22$.
    Morphologically selected stars are then those objects with
    sizes in each passband smaller than the PSF size plus $(0\farcs3,0\farcs4)$ in the $(g,i)$ bands for the \citet{Okamoto2015} data set, and $(0\farcs3,0\farcs2,0\farcs3)$ in $(g,r,i)$ for the \citet{Smercina2020} data set. In artificial galaxy detection experiments, these thresholds yielded the faintest detection limits that could be achieved with modest contamination. 
  \item Stellar Locus: For the \citet{Smercina2020} dataset, the
    three-passband coverage allows one to further select sources to
    have $g-r$ and $r-i$ colors similar to stars. This selection is
    described in more detail in \citet{Smercina2020}.
  \item Nearest Neighbor: The \citet{Smercina2020} data are of
    uniform enough quality that supervised machine-learning techniques
    can be used. Nine quantities are
    used: $i$-band magnitude, $g-r$, $r-i$, and the object size in the R.A.\ and decl.\ directions for each
    of $g$, $r$ and $i$ bands. Each quantity was normalized to have
    an outlier-resistant standard deviation of unity to carry similar
    weights in the classification. 
    We select a training set of $\sim26,000$ background objects in areas distant from the M81 group galaxies. In order to assemble a training set
    of likely stars, we statistically subtract these background
    objects from the population of objects in a star-rich region of equal area in the
    stellar envelopes of M82 and NGC 3077. For each background object,
    the nearest match in the star-rich region is identified in
    9-dimensional space, and this object is discarded from the
    dataset, leaving a sample that is likely to
    contain primarily stars ($\sim13,000$
    objects).
    These star and background training sets
    then classify all objects using the
    majority vote of the 11 nearest neighbors in this 9-dimensional
    space (using \texttt{scikit.learn.neighbors}). Classifications for the background and star-rich regions
    were generated using alternate regions. The detailed
    choice of training regions does not affect any of our conclusions. 
\end{enumerate}

\begin{figure*}[ht!]
\centering
\includegraphics[width=0.85\textwidth]{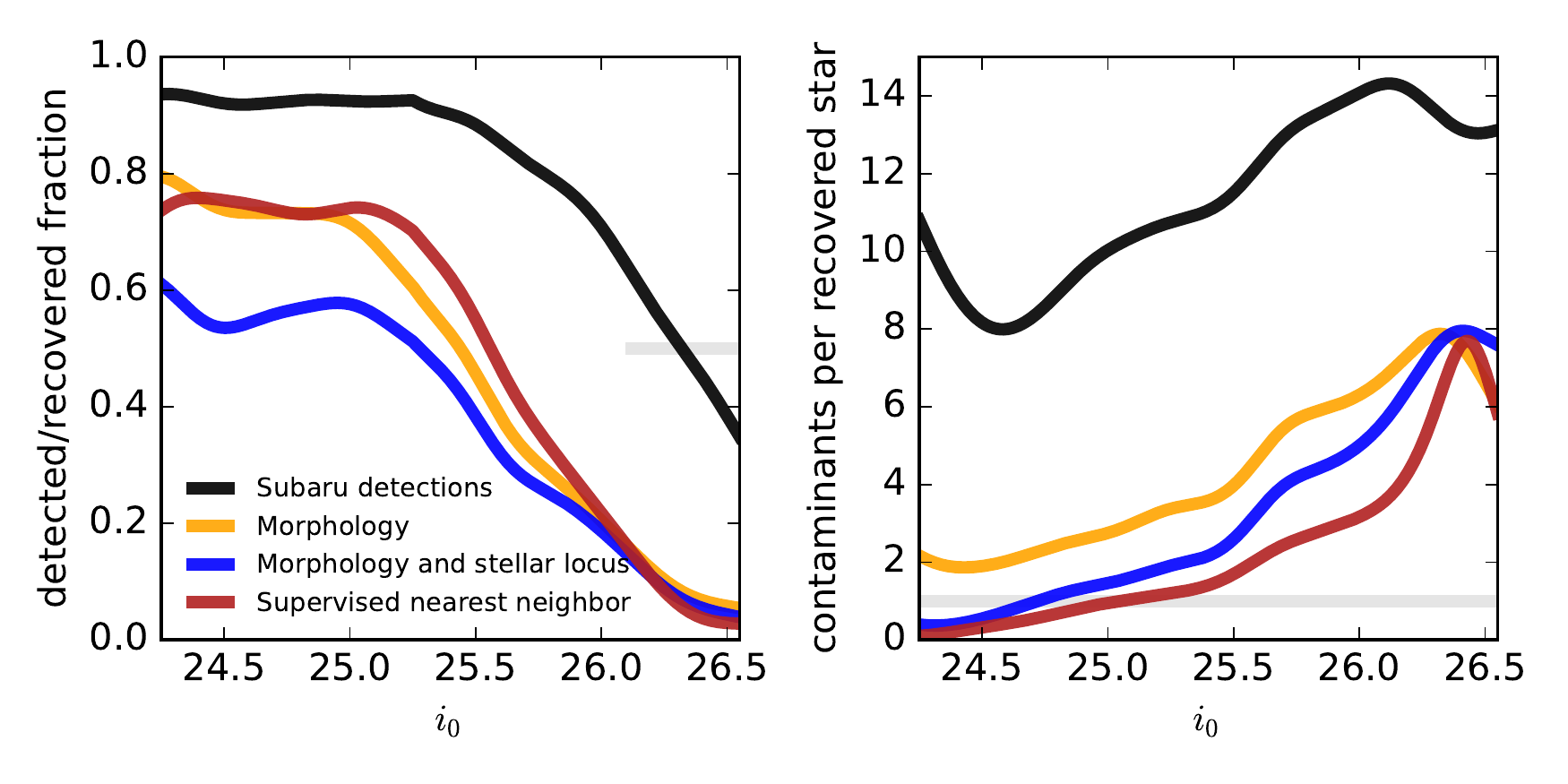}
\caption{Left: completeness, quantified using Gaussian kernel density estimation with $\sigma = 0.15$\,mag, as measured using M81
  halo stars with HST imaging. While most HST stars have a counterpart
  in the Subaru catalog (black line), star--galaxy separation
  techniques are so selective that they discard many real stars up to 1 mag
  brighter than the nominal 50\% completeness limit of $i\sim 26.3$ (gray line). Right:
  contamination by galaxies from each selection. The gray line shows the case when the number of contaminants equals the number of recovered stars. Galaxies
  dramatically outnumber stars in M81's halo; star--galaxy separation
  cuts down the contamination considerably.
  \label{fig:compcont}}
\end{figure*}

We illustrate the performance of the three different star--galaxy
separation techniques in Fig.\ \ref{fig:compcont}. We use as ground
truth stars in uncrowded regions of M81's halo
with Hubble Space Telescope (HST) imaging from the GHOSTS survey
\citep{RS11}\footnote{The relatively small
  number of HST stars allows star--galaxy separation testing
  but was insufficient to act as a training set for star--galaxy
  separation.}. The overall completeness
remains above 90\% for $i<25.4$, dropping rapidly thereafter (black curve; left panel). The right panels show the contamination $\Sigma_{contamination}$ divided by the number of detected stars, which is the position-dependent density of stars $\Sigma_{stars} (\alpha,\delta)$ multiplied by the completeness $c_{stars}$ (we analyze uncrowded regions and can neglect spatial variations in completeness; \citealt{Smercina2020}). The GHOSTS fields used to quantify contamination have low stellar density and so give contamination measures $\frac{\Sigma_{contamination}}{\Sigma_{stars} (\alpha,\delta) c_{stars}}$ that are higher than would be expected at the positions of our UFD candidates, but give a robust measure of the {\it relative} performance of star--galaxy separation methods. Notwithstanding this limitation, it is clear that star--galaxy separation is crucial --- in the GHOSTS fields, 
there are 8-14$\times$ more galaxies than stars at such limits (black curve, right panel),
motivating stringent star--galaxy separation. The three star--galaxy selections 
already lower completeness at $i<25.5$ and strongly reduce it for $i>25.5$; contamination by background galaxies is, however, dramatically reduced, permitting a search for UFD candidates that will be less overwhelmed by the clustering of background compact sources.

\section{UFD candidate identification}
\label{sec:cand_sel}

\begin{figure*}
\gridline{\fig{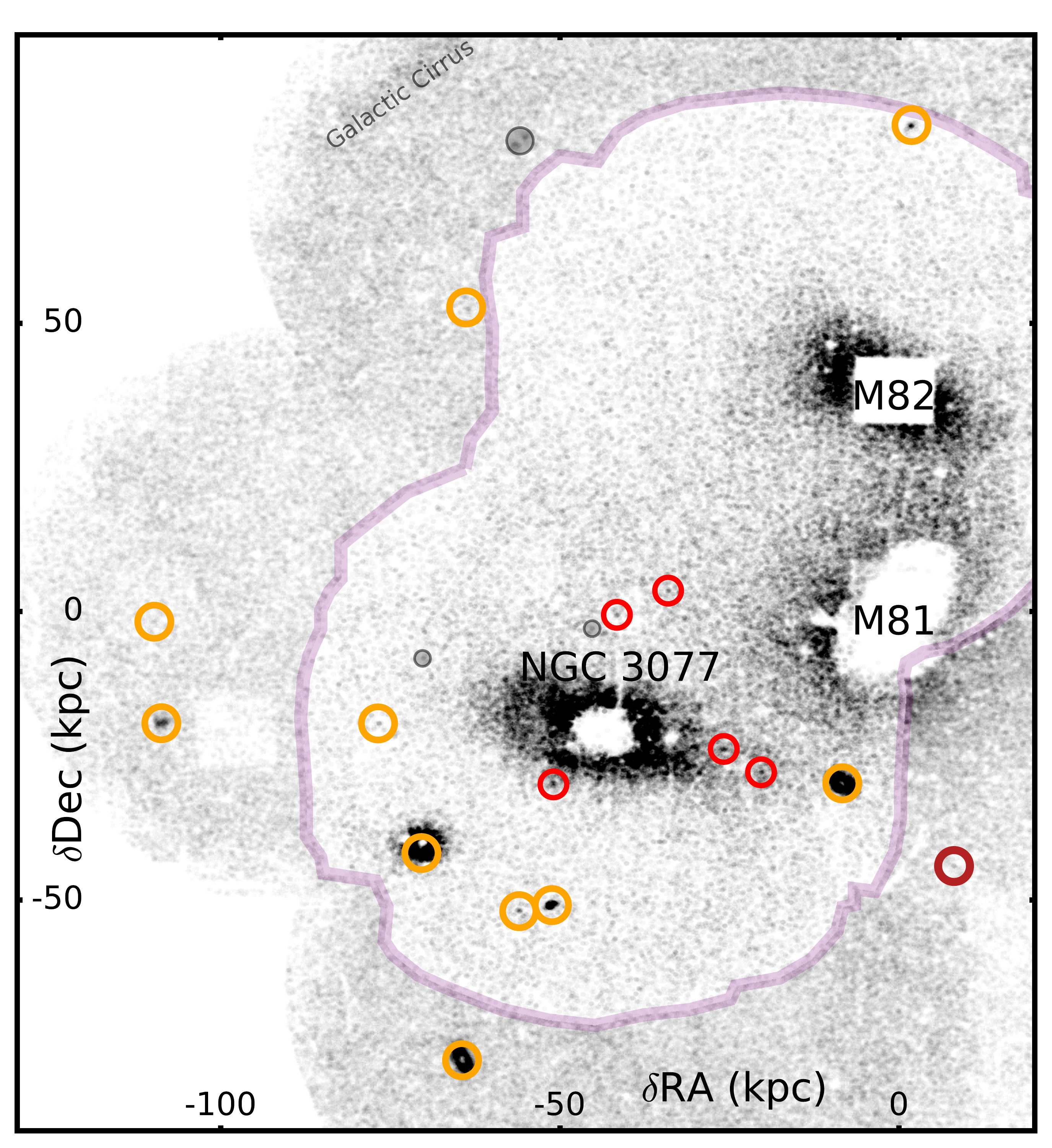}{0.5\textwidth}{}
          \fig{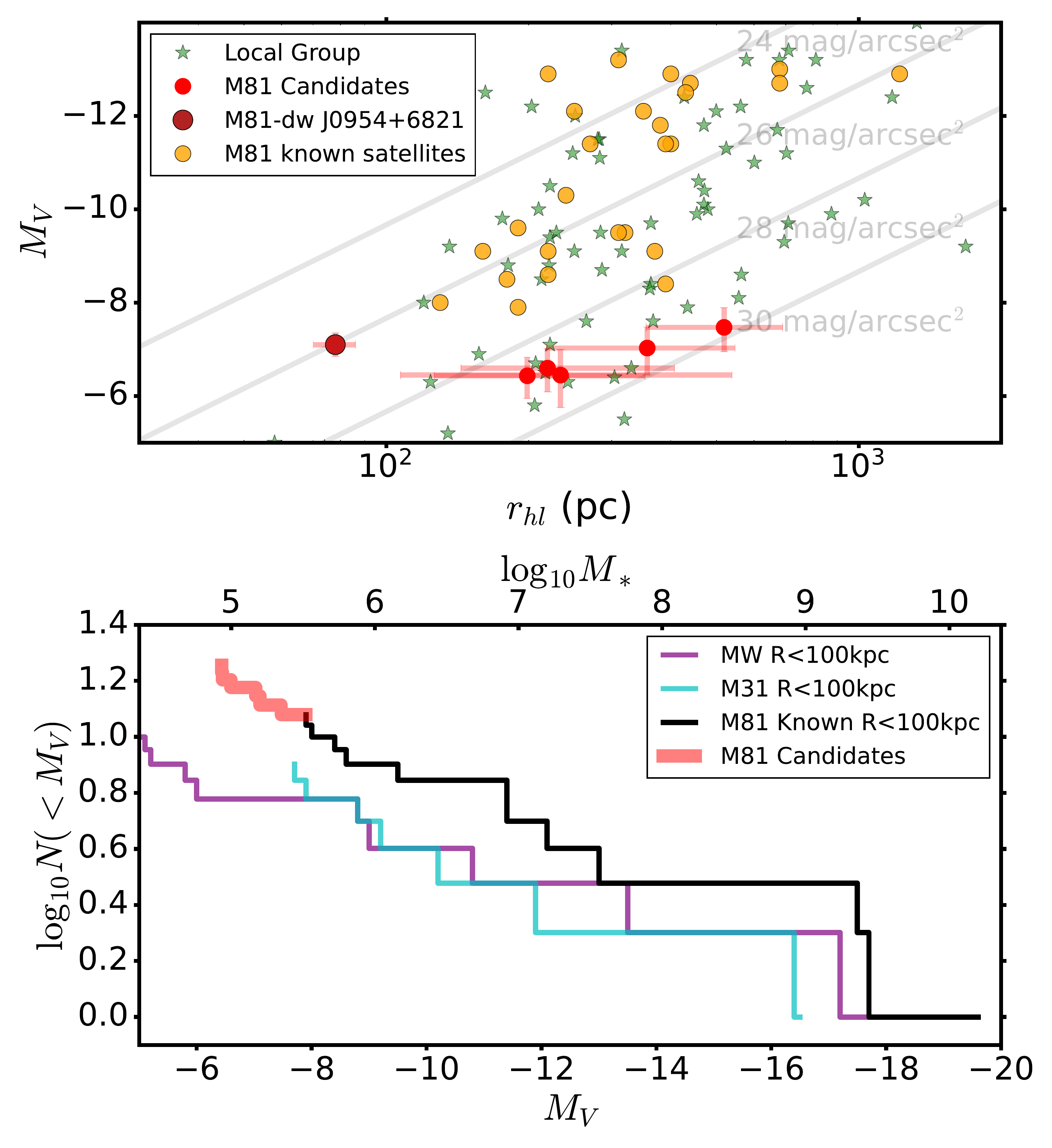}{0.5\textwidth}{}
          }
\caption{Left: the distribution of known galaxies (orange) and candidate UFDs (red) in the M81 group. The KDE map with a 400\,pc kernel is shown; the $x-$ and $y-$axes are the projected distance at the distance of the M81 group in the R.A.\ and decl.\ direction in kpc. The outer parts are from \citet{Okamoto2015}; inside the purple outline the $gri$ \citet{Smercina2020} dataset is used. The definite dwarf M81-dw J0954+6821 is in brick red. The large gray circle shows the location of background galaxy UGC 5423; small gray circles show rejected overdensities. Top right: the magnitude--size relation for Local Group galaxies (green stars), known M81 satellites (orange), the new M81 definite UFD M81-dw J0954+6821 (brick red), and M81 UFD candidates (red). Lines of constant enclosed surface brightness within the half-light radius are shown in gray. Bottom right: the luminosity function of Milky Way, M31, and M81 galaxies within $D<100$\,kpc for the Milky Way (purple), and $R_{proj} < 100$\,kpc for M31 (cyan) and M81's known satellites (black) and candidates$+$new UFD (red).
  \label{fig:cand_sel}}
\end{figure*}

Candidate UFDs will appear as overdensities of objects with
the colors and magnitudes expected for metal-poor red giant branch (RGB) stars \citep[e.g.,][]{Martin2008}. We select stars
within the $(g-i,i)$ polygon with corners $(0.75, 26.5), (1.45, 26.5), (1.85, 24.2)$, and $(1.4, 24.2)$\footnote{This selection region encloses the metal-poor RGB population characteristic of M81's outer stellar halo (see the 25--35\,kpc and 35--45\,kpc radial bins in Fig.\ 7 of \citealt{Smercina2020}).}. This color--magnitude cut is sufficiently red to avoid contamination by the young stars that are widespread across the whole M81 group \citep[e.g.,][]{Okamoto2015}. In order to identify overdensities with sizes comparable to the half-light radii of $M_V \sim -7$ UFDs in the Local Group, we determine the density of metal-poor RGB stars using kernel density estimation using a top-hat kernel with radii of 200 and 400\,pc; we
sample the distribution on 100\,pc scales\footnote{Tests show that these kernels recover artificial galaxies with properties similar to Local Group UFDs.}. We demand that an overdensity has a Poisson probability of being drawn from the spatially varying background (assessed using a 4\,kpc top-hat radius) of $P<10^{-6}$.
Rapid changes in density in the inner parts of bright galaxies yield spurious overdensities; we therefore conservatively exclude any recovered overdensity within 22, 12, 10, 5, and 4 projected kpc from M81, M82, NGC 3077, IKN, and
KDG61, respectively. We then determine a more tailored
measure of significance by allowing modest shifts in the center and
choosing the best significance in a range of apertures between 100
and 800\,pc. Candidates are those
objects that have final probability $P<10^{-7}$ of being drawn from the
background by chance alone. We search each dataset separately: \citet{Okamoto2015} using the Morphology cut, and using all three star--galaxy separation methods in the \citet{Smercina2020} dataset. In areas where both datasets overlap, we choose only candidates detected in the \citet{Smercina2020} dataset. Given the search area, probing $\sim 2\times10^5$ independent 200\,pc radius apertures,  $\sim 0.02$ candidates would be
expected from chance alone.

We quantify the expected degree of contamination by artifacts or groups of background galaxies by analyzing $gri$ archival data taken by the Subaru Strategic Program of a four-pointing deep HSC mosaic in the COSMOS field \citep{Aihara2022}, where the probability of finding a real Local Volume UFD is very low. This has similar area and depth but has slightly worse seeing. The morphological (stellar locus) cut gives three (one) candidates across the four HSC fields with $P<10^{-7}$. 

In the M81 datasets, we recover all known dwarf galaxies in the search area (Fig.\ \ref{fig:cand_sel}, left) in addition to a background galaxy UGC 5423 ($D \sim 9$\,Mpc; large gray circle). We find 11 additional candidates in the seven-pointing \citet{Okamoto2015} mosaic using the morphology-only cut ($\sim6$ would be expected from our analysis of COSMOS). After experimentation, we found no robust algorithmic methods for rejecting clumps of background objects or spurious detections, so we visually inspect each candidate. Eight candidates are clearly spurious (artifacts near bright stars, field edges, or galactic cirrus) and are discarded completely. Two overlap with the deep coverage with more stringent star--galaxy separation, and are vetoed by that deeper dataset. One candidate --- M81-dw J0954+6821 --- is compact enough to show diffuse surface brightness; furthermore, the diffuse brightness is bluer in color than the resolved RGB stars, as is expected for a partially resolved dwarf galaxy (as the diffuse light is dominated by bluer subgiants and main-sequence turnoff stars; see e.g., Figure 1 of \citealt{Sand2022}). On this basis, we argue that it is a clear dwarf galaxy (brick red circle in
Fig.\ \ref{fig:cand_sel}; Table \ref{Table}). 

\begin{figure*}
\centering
\includegraphics[width=0.6\textwidth]{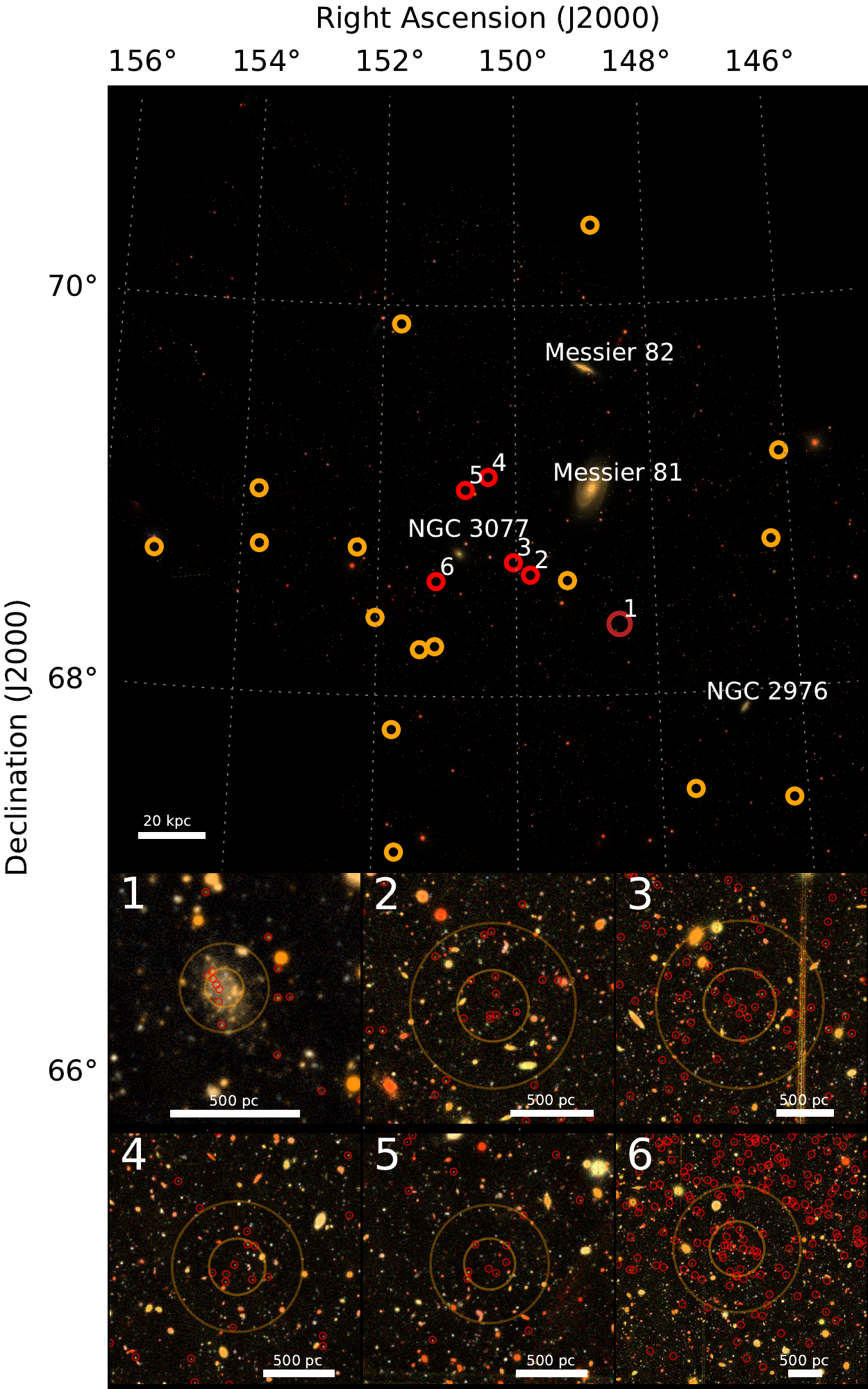}
\caption{The distribution of known dwarf galaxies (orange) and our sample (brick red shows the definite dwarf M81-dw J0954+6821, labeled 1; red shows the other candidates, labeled 2--6). The background is the SDSS $gri$ mosaic of the M81 region. We also show $gri$ postage stamps of the candidates. Red circles show likely metal-poor RGB stars in the postage stamps, and the orange circles show the 50\% and 90\% light radii. 
\label{fig:postage_stamps}}
\end{figure*}

\begin{figure*}
\plotone{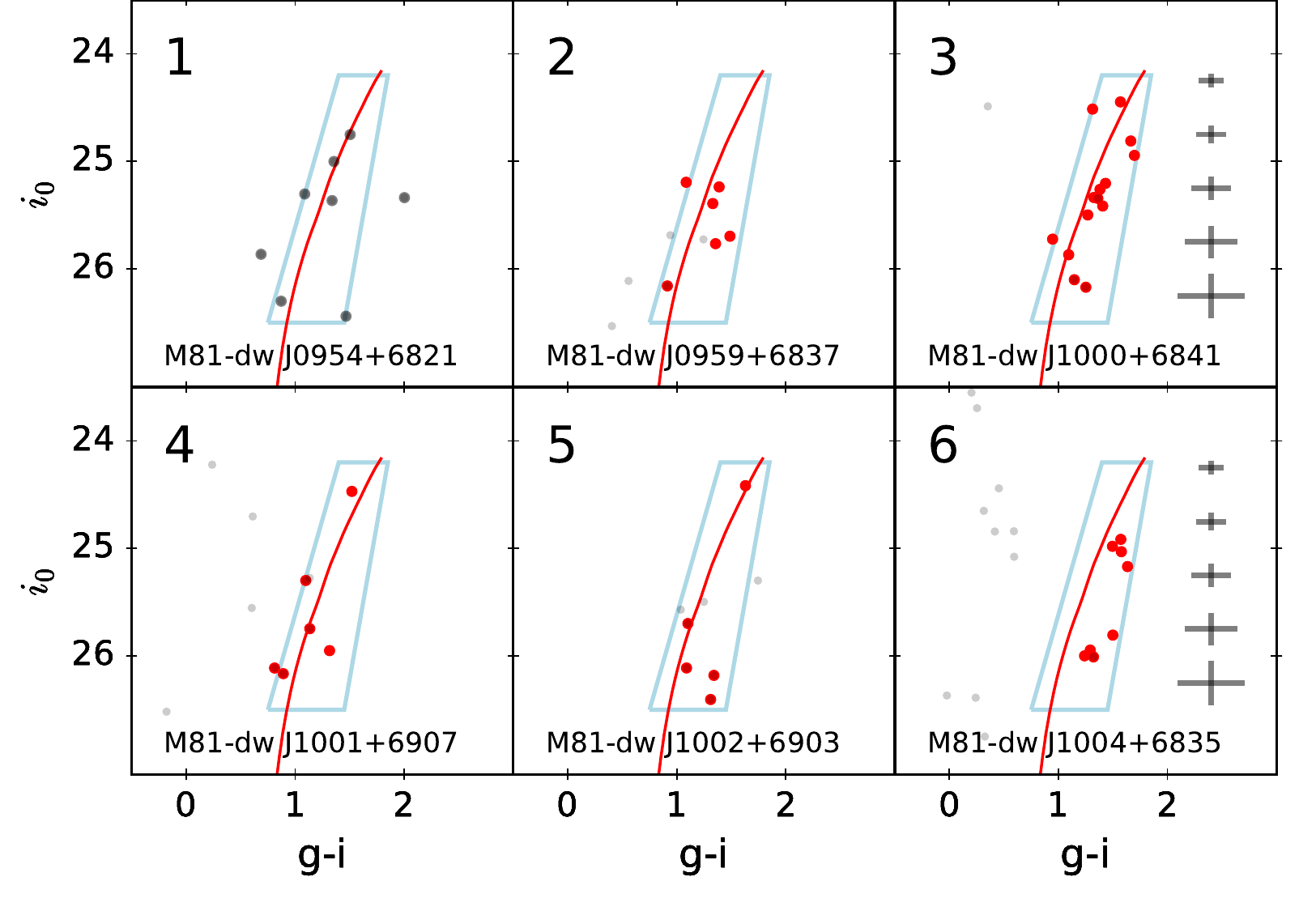}
\caption{Background-subtracted CMDs within the 80\% light radius. Morphologically selected stars are shown in gray; in candidates 2--6 with $gri$ coverage, stellar-locus-selected stars with colors consistent with a metal-poor RGB are shown in red. Representative error bars are shown in the rightmost panels. A 13.1\,Gyr ${\rm[M/H]}=-1.6$ isochrone is shown in red, and the selection region for RGB stars is shown in blue. In all cases, objects with close color--magnitude matches in a background annulus of equal area were subtracted from the CMD. 
\label{fig:CMDs}}
\end{figure*}

We find eight candidates in the two-pointing deep dataset with $\ge5$ stars after background subtraction, where we would expect only 0.5 spurious candidates from our COSMOS analysis. One candidate is close to a bright star and is rejected outright. Seven candidates remain (Table \ref{Table}). One of them is close enough to a bright background galaxy that we are concerned that the point sources might be globular clusters around those nearby galaxies (an important contaminant of candidate M31 UFDs;  \citealt{Martin2013}). Another appears to be a background galaxy group, given a clear concentration of blue compact sources in that candidate's color--magnitude diagram (CMD). The remaining five appear to be stellar in
nature, and we retain them as candidates, shown as red circles in
Fig.\ \ref{fig:cand_sel} (see Table \ref{Table}). We show postage stamps of the candidates in Fig.\ \ref{fig:postage_stamps}. Fig.\ \ref{fig:CMDs} shows background-corrected CMDs for each candidate. We show objects within the 80\% light radius (determined from the fits described in \S \ref{sec:prop_dist}). In order to background-subtract the CMD, we choose a background annulus of equal area (with inner radius $3.5\,r_{e}$), and for every object in that background annulus we choose the closest match in color--magnitude space of objects within the 80\% light radius and discard it from the CMD. Only the remaining objects --- those that are in excess of the background in that region --- are shown in Fig.\ \ref{fig:CMDs}. 
Gray symbols show background-corrected morphologically selected `stars'; red symbols (in panels 2--6) show stellar-locus-selected stars\footnote{The blue `stars' in M81-dw J1004+6835's CMD are likely to be real, from young stars in M81's {\sc Hi} tidal field \citep{Okamoto2015}.}.

\newcolumntype{s}{!{\extracolsep{-4pt}}l!{\extracolsep{0pt}}}
\newcolumntype{p}{!{\extracolsep{-4pt}}c!{\extracolsep{0pt}}}
\begin{deluxetable*}{spssssss}
\tablecaption{Dwarf galaxy candidates and likely contaminants in the M81 group\label{Table}}
\tablewidth{0pt}
\tablehead{
\colhead{Name} & \colhead{Number} & \colhead{R.A.\ (J2000)} & \colhead{Decl.\ (J2000)} & \colhead{$M_V$} & \colhead{$r_e$} & \colhead{Data Set} & \colhead{Note} \\ 
\colhead{} & \colhead{} &\colhead{(deg)} & \colhead{(deg)} & \colhead{(mag)} & \colhead{(pc)} & \colhead{} & \colhead{}
}
\startdata
M81-dw J0954+6821$^a$ & 1 & $148.5292\pm0.0008$ & $68.3641 \pm 0.0003$ & $-7.1\pm0.25$ & 
$78\pm8$ & Okamoto & Definite dwarf \\
M81-dw J0959+6837 & 2 & $149.7931\pm0.0037$ & $68.6212_{-0.0013}^{+0.0022}$ & $-6.5_{-0.6}^{+0.7}$ & $230_{-130}^{+310}$ & Smercina &  \\ 
M81-dw J1000+6841 & 3 & $150.0402_{-0.0055}^{+0.0070}$ & $68.6855_{-0.0014}^{+0.0017}$ & $-7.0_{-0.5}^{+0.6}$ &  $360_{-140}^{+190}$ &  Smercina &  \\
M81-dw J1001+6907 & 4 & $150.4039\pm0.0037$ & $69.1224\pm0.0014$ & $-6.6_{-0.4}^{+0.5}$ & $220_{-80}^{+190}$ & Smercina &  \\ 
M81-dw J1002+6903 & 5 & $150.7405\pm0.0035$ & $69.0559_{-0.001}^{+0.0014}$ & $-6.4_{-0.4}^{+0.5}$ & $200_{-70}^{+150}$ & Smercina &  \\ 
M81-dw J1004+6835 & 6 & $151.1613\pm0.0071$ & $68.5916\pm0.0024$ & $-7.5_{-0.4}^{+0.5}$ & $520\pm160$ & Smercina & Superimposed on \\
& & & & & & & NGC 3077 tidal debris \\ 
\hline
M81-dw J1008+6856 & \nodata & $152.0104_{-0.0080}^{+0.0033}$ & $68.9367_{-0.0019}^{+0.0022}$ & $-6.5_{-0.5}^{+0.8}$ &  $470_{-410}^{+230}$ & Smercina & Likely Background \\
& & & & & & & Galaxy Cluster \\ 
M81-dw J1003+6901 & \nodata & $150.9062\pm0.0038$ & $69.0187\pm0.0018$ & $-6.2_{-0.5}^{+0.6}$ & $220_{-100}^{+270}$ & Smercina & Possible background GCs  \\ 
\enddata
\tablecomments{$^a$ M81-dw J0954+6821 has high enough surface brightness to have a well-measured ellipticity $b/a \sim 0.55\pm0.1$ and PA$\sim 25\pm8$.
}
\end{deluxetable*}

\section{The properties and distribution of dwarf and ultrafaint dwarf candidates}
\label{sec:prop_dist}

\subsection{Candidate properties}

In order to estimate candidate properties, we assume that each is at the M81 group distance of $D=3.6$\,Mpc.
We follow \citet{Martin2008} in fitting a two-dimensional exponential profile with uniform background to the distribution of detected stars. The candidates have so few stars that their ellipticity and position angle are virtually unconstrained. We therefore instead fit a four-parameter model: R.A., decl., the major-axis half-light radius $r_e$, and the number of stars.
A Markov Chain Monte Carlo maximum likelihood fit is performed using \texttt{emcee} \citep{ForemanMackey2013}, using uniform priors in all parameters except position; a Gaussian prior on position is applied with $\sigma = 400$\,pc in each direction. We estimate absolute magnitude by scaling the observed number of stars by the magnitude-dependent completeness (Fig.\ \ref{fig:compcont}; this correction is roughly a factor of two) and the absolute magnitude per detected RGB star, estimated from a 13.3\,Gyr old [M/H]$=-1.6$ Padova isochrone\footnote{\url{http://stev.oapd.inaf.it/cmd}, using the PARSEC evolutionary tracks version 1.2S \citep{Bressan2012}.}  (given the ages and metallicities of similarly luminous UFDs; \citealt{Brown2014}, \citealt{Simon2019}; shown in Fig.\ \ref{fig:CMDs}). Dwarf galaxy M81-dw J0954+6821 has high surface brightness and is crowded; we therefore directly calculate the flux and half-light radius from the image itself. The results of these fits are given in Table \ref{Table}. 

M81-dw J0954+6821 has a high surface brightness within the half-light radius ($\langle \mu_V \rangle_{<r_e} \sim $26 mag\,arcsec$^{-2}$; Fig.\ \ref{fig:cand_sel}a) and is a definite dwarf galaxy. It has well-measured parameters: $r_e=78\pm8$pc, $b/a = 0.5\pm0.1$, PA$=25\deg$, and $M_V = -7.1\pm0.25$. Its magnitude error is dominated by background uncertainty. Its CMD (Fig.\ \ref{fig:CMDs}) is sparse, due at least partially to crowding. In contrast, our most diffuse ($\langle \mu_V \rangle_{<r_e} \sim $29.5 mag\,arcsec$^{-2}$) candidate is M81-dw J1004+6835 --- it is just $\sim 10\,$kpc south of NGC 3077 and is superimposed on a rich stellar population from NGC 3077 itself. The other four candidates have $M_V$ between $-6.5$ and $-7$ and  $r_e$ values between 200 and 350\,pc. Due to their extreme surface brightnesses ($\langle \mu_V \rangle_{<r_e} \sim $29 mag\,arcsec$^{-2}$), no diffuse light can be detected; deeper CMD data from HST or JWST will be required for confirmation. 

It is clear why these candidates have so far evaded detection: all candidates are fainter than all known M81 group dwarf galaxies, and all but one of the candidates have much lower surface brightness than known M81 dwarfs. Yet the M81 satellites' properties are not unexpected, overlapping with the ranges of magnitudes and sizes of dwarf galaxies and UFDs in the Local Group (Fig.\ \ref{fig:cand_sel}, top right). 

Within or near the Local Group, there are two analogs to the relatively compact galaxy  M81-dw J0954+6821: Pegasus V/Andromeda XXXIV ($D=690$\,kpc; \citealt{Collins2022}) and Tucana B ($D=1.4$\,Mpc; \citealt{Sand2022}). In their relatively high luminosities and large sizes, M81-dw J1004+6835 and M81-dw J1000+6841 are most similar to Canes Venatici I/Andromeda IX and Hercules/Andromeda XXIV, respectively. The remaining dwarfs are analogs of Bo\"otes I, or equivalently Andromeda XIII or XXII. 

\subsection{The spatial distribution of candidates}

Fig.\ \ref{fig:cand_sel} shows that candidates are not distributed uniformly in the M81 group. One might have expected the M81 group satellites to be clustered around M81 itself, or potentially, given the evidence for \edit1{satellite} infall with the Magellanic Clouds, around M82, M81's largest satellite. Yet, instead, they are clustered around NGC 3077. One immediate implication is that most of these candidates are likely to be real --- our COSMOS blank-field study shows that spurious candidates are more spatially uniform. 

Given the spatially varying seeing, fully accounting for completeness requires forward modeling given expected satellite distributions and is beyond the scope of this work. The \citet{Smercina2020} dataset has uniform depth, permitting instead a preliminary estimate of significance.  In the northern field, excluding M82, there is one known fainter satellite. In NGC 3077's field, excluding NGC 3077, there are five known fainter satellites and five new candidates (10 total). The chance of drawing 10 satellites from a Poisson distribution if the mean is 1 is $\sim 10^{-8}$; alternatively, the chance of drawing one satellite if the mean is 10 is $\sim 5 \times 10^{-4}$. If the mean is 5.5 (the average), the chance of drawing one or less for one draw and 10 or more for the other is $\sim 0.027 \times 0.025$, or $7 \times 10^{-4}$. We conclude that there is less than a $\sim 7 \times 10^{-4}$ chance that this difference in satellite counts is from chance alone\footnote{Including satellites or candidates outside the \citet{Smercina2020} footprint, excluding those galaxies that are closer to NGC 2976 with R.A.$<148^{\circ}$ or decl.$<68^{\circ}$ (2 vs.\ 14), $P<2 \times 10^{-4}$ instead. Restricting our attention conservatively to clear dwarf galaxies in this area (2 vs.\ 9) gives $P<0.5\%$.}.

\section{Discussion} \label{sec:disc}

Assuming that these candidates are real, and neglecting completeness corrections, we illustrate the impact of these new discoveries on the M81 group luminosity function within a projected radius of $R_{proj} < 100$\,kpc (Fig.\ \ref{fig:cand_sel}, bottom right), in comparison with $D < 100$\,kpc Milky Way satellites (purple; \citealt{DW2020}) and $R_{proj} < 100$\,kpc M31 satellites (cyan; \citealt{McConnachie2018}). This satellite and these candidates extend the M81 group luminosity function faintward by 1.5 mag (or a factor of 4 in luminosity). Despite M81's stellar mass being comparable to those of the Milky Way or M31 (with $M_*/10^{10} M_{\odot} \sim 6$, 6 and 10 respectively; \citealt{Bell2017}), the M81 group is richer than either the Milky Way's or M31's within equivalent radii. At this stage, we refrain from ascribing this difference to a single cause, as several factors should (or have been observed to) correlate with the overall number of satellites --- e.g., the stellar mass of the host, virial mass of the DM halo, and delivery of satellites by recent group accretions \citep{Carlsten2021,DSouza2021,Smercina2022}.   

One of the most interesting features of these candidates is that all of them are projected close to NGC 3077. Clearly, confirmation of the candidates via deep high-resolution HST or JWST photometry and velocity measurements (via semi-resolved spectroscopy, following, e.g., \citealt{Toloba2016}) would be very valuable to understand this association with NGC 3077. This strengthens the recognition by \citet{Chiboucas2013} of the M81 group's highly flattened satellite system and is reminiscent  of M31's asymmetric satellite system, where 80\% of its satellites are preferentially on the near side of M31 (Fig.\ 13 in \citealt{Savino2022}). 

The spatial coincidence of NGC 3077 and most of the M81 group satellites is surprising. Since satellite number should scale with DM halo mass (e.g., \citealt{Jiang2015}), M81 should host most of the satellites. Yet the satellite distribution is extremely asymmetric, indicating that many of the M81 group's satellites were recently accreted as a group and have not yet had time to phase-mix (as discussed by, e.g., \citealt{DSouza2021}). This work adds further evidence that satellites of satellites are important in building up the satellite populations of Milky Way--mass galaxies (see also, e.g., \citealt{LiHelmi2008}, \citealt{Deason2015}, \citealt{Patel2020}, \citealt{DSouza2021}). 

In this picture, one would expect most of the recently arrived satellites to be associated with M82, which is clearly undergoing tidal disruption \citep{Okamoto2015,Smercina2020} and has a stellar mass 10$\times$ larger than the next most massive satellite, NGC 3077  \citep{Smercina2020}. M82 should then have had a more massive DM halo, and therefore have delivered a substantial number of satellites \citep{Smercina2022}. Yet the satellites are spatially clustered around NGC 3077. It is possible that these satellites were instead delivered by NGC 3077, and for some reason M82 had few satellites.
Yet it is also possible that these satellites and NGC 3077 itself were M82's satellites.  Satellites are stripped relatively early as a group falls into a larger potential well, while M82, due to its much larger mass, may be subject to much stronger dynamical friction \citep{DSouza2021}. 
It is therefore possible that these satellites were in fact previously all part of M82's group but were `left behind' by M82, as it loses energy through dynamical friction as it merges with M81. 

\section{Conclusions} \label{sec:conc}

In this letter, we report the discovery using resolved-star techniques of one new M81 group UFD (similar to Tucana B) and present five lower surface brightness candidate UFDs (similar to Canes Venatici I, Hercules, and Bo\"otes I), with absolute magnitudes reaching toward $M_V \sim -6$. While these candidates, with $\langle \mu_V \rangle_{<r_e}$ typically between 28 and 29.5 mag\,arcsec$^{-2}$, require HST or JWST follow-up for confirmation, blank-field searches with comparable areas yield $<1$ candidate, and the properties of these candidates overlap with those of Local Group UFDs, suggesting that most or all of the candidates should be real. 
The candidates are not distributed uniformly but instead cluster strongly around NGC 3077 --- the third-brightest galaxy in the central parts of the M81 group --- at the $<$99.9\% significance level. This underlines the importance of group accretion in shaping the satellite populations of nearby galaxies. However, it also raises a puzzle of why M81 and M82 --- both more massive in stars, and likely more massive in DM --- do not host more satellites; this puzzle remains unresolved. 

\begin{acknowledgments}
This work was partly supported by the National Science Foundation through grant NSF-AST 2007065 and by the WFIRST Infrared Nearby Galaxies Survey (WINGS) collaboration through NASA grant NNG16PJ28C through subcontract from the University of Washington. AM gratefully acknowledges support by FONDECYT Regular grant 1212046 and by the ANID BASAL project FB210003, as well as funding from the Max Planck Society through a “PartnerGroup” grant. This research has made use of NASA's Astrophysics Data System Bibliographic Services. 

Based on observations utilizing Pan-STARRS1 Survey. The Pan-STARRS1 Surveys (PS1) and the PS1 public science archive have been made possible through contributions by the Institute for Astronomy, the University of Hawaii, the Pan-STARRS Project Office, the Max-Planck Society and its participating institutes, the Max Planck Institute for Astronomy,
Heidelberg and the Max Planck Institute for Extraterrestrial Physics, Garching, Johns Hopkins University, Durham University, the University of Edinburgh, the Queen’s University Belfast, the Harvard--Smithsonian Center for Astrophysics, the Las Cumbres Observatory Global Telescope Network Incorporated, the National Central University of Taiwan, the Space Telescope Science Institute, the National Aeronautics and Space Administration under grant No. NNX08AR22G issued through the Planetary Science Division of the NASA Science Mission Directorate, the National Science Foundation grant No. AST-1238877, the University of Maryland, Eotvos Lorand University (ELTE), the Los Alamos National Laboratory, and the Gordon and Betty Moore Foundation. 

Based on observations obtained at the Subaru Observatory, which is operated by the National Astronomical Observatory of Japan, via the Gemini/Subaru Time Exchange Program. We thank the Subaru support staff for invaluable help preparing and carrying out the observing run. 

The authors wish to recognize and acknowledge the very significant cultural role and reverence that the summit of Maunakea has always had within the indigenous Hawaiian community. We are most fortunate to have the opportunity to conduct observations from this mountain.

\software{\texttt{HSC Pipeline} \citep{Bosch2018}, \texttt{Matplotlib} \citep{matplotlib}, \texttt{NumPy} \citep{numpy}, \texttt{Astropy} \citep{astropy}, \texttt{SciPy} \citep{scipy}, \texttt{Scikit-learn} \citep{scikit-learn}}

\end{acknowledgments}


\bibliographystyle{aasjournal}



\end{document}